\RequirePackage{fix-cm}
\documentclass[smallcondensed]{svjour3}
\usepackage[utf8]{inputenc}
\usepackage[colorlinks=true,linkcolor=blue,citecolor=blue,urlcolor=blue]{hyperref}
\usepackage{mathptmx}
\usepackage{listings}
\usepackage{lstlangcoq}
\lstnewenvironment{coq}
  {\lstset{language=Coq,basicstyle=\sffamily\small,mathescape=true,columns=fullflexible}}
  {}
\newcommand{\lst}[1]{\hbox{\lstinline{#1}}}

\journalname{Journal of Automated Reasoning}
\title{Efficient Extensional Binary Tries
\thanks{
There are no conflicts of interest.  Code is available at
\url{https://github.com/xavierleroy/canonical-binary-tries/tree/v2}.
}
\thanks{
This version of the article has been accepted for publication,
  after peer review, and is subject to Springer Nature’s AM terms of use, but is not the Version of Record and does not reflect post-acceptance improvements, or any corrections. The Version of Record is available online at: \url{https://doi.org/10.1007/s10817-022-09655-x}
}
}

\titlerunning{Efficient Extensional Binary Tries}

\author{Andrew W.~Appel
  \and Xavier Leroy
}
\authorrunning{A.~W.~Appel and X.~Leroy}
\institute{Andrew W.~Appel \at
                Department of Computer Science, Princeton University, 35 Olden Street, Princeton, NJ, 08540           \\
                \email{appel@princeton.edu}
                \and
          Xavier Leroy \at
          Coll\`ege de France, PSL University, 3 rue d'Ulm, 75005 Paris \\
          \email{xavier.leroy@college-de-france.fr}
}

\date{}

\begin{document}

\maketitle

\begin{abstract}
Lookup tables (finite maps) are a ubiquitous data structure.  In pure
functional languages they are best represented using trees instead of
hash tables.  In pure functional languages within constructive logic,
without a primitive integer type, they are well represented using
binary tries instead of search trees.

In this work, we introduce \emph{canonical binary tries}, an improved
binary-trie data structure that enjoys a natural extensionality
property, quite useful in proofs, and supports sparseness more
efficiently.  We provide full proofs of correctness in Coq.

We provide microbenchmark measurements of canonical binary tries
versus several other data structures for finite maps,
in a variety of application
contexts; as well as measurement of
canonical versus original tries in two big, real systems.
The application context of \emph{data structures contained
in theorem statements} imposes unusual requirements
for which canonical tries are particularly well suited.
\end{abstract}

\section{Introduction}

Lookup tables---finite maps from identifiers to bindings---are a central data structure in many kinds of programs.  
We are particularly interested in programs that are proved 
correct---compilers, static analyzers, program verifiers.
Those programs are often written in pure functional languages,
since the proof theory of functional programming is more tractable
than those of imperative or object-oriented languages.
Thus we focus on applications written in the functional languages
internal to the logics of Coq or HOL.
Such programs \emph{may} be ``extracted'' to programming languages
such as OCaml, Standard ML, or Haskell, and compiled with optimizing
compilers for those languages; the CompCert C compiler \cite{Leroy-CompCert-CACM}
and the Verasco static analyzer \cite{Jourdan-LBLP-2015} are examples.
On the other hand, some programs, such as the
Verified Software Toolchain program verifier \cite{appel11:esop},
don't use extraction:
those programs are reduced within the logic
of a proof assistant, so that the program verifier can gracefully
interact with user-driven interactive proof in the ambient logic.

Many proved-correct programs of this kind deliberately use an
impoverished set of primitives, so that fewer axioms need to be
assumed (or equivalently, fewer lines of proof-checker kernel code need to be
trusted).  In particular, there may be no primitive integer type
implemented in a machine word.  Instead, integers (or whatever type is
used for \emph{keys,} the domain of the finite maps) may be
constructed from the inductive data types of the logic.
Consequently, we cannot assume that a less-than comparison of two
keys takes constant time; it will be logarithmic in
the magnitude of the numbers.

These criteria led to a preliminary design for a data
structure and its algebraic interface: binary tries, also called
positive trees (for reasons that will become clear).  This data
structure was used in the CompCert C compiler,
the Verified Software Toolchain (VST),
the Verasco verified static analyzer, and in other applications.
In the process, we have learned to demand more and more
criteria of efficiency and provability from them:
\begin{enumerate}
\item[] \hspace*{-1em}\emph{Core Requirements}
\item Basic operators: \lstinline{empty} (the finite map with an empty domain),
  \lstinline{get} (the lookup operator), and \lstinline{set} (the update operator).
\item Proofs\footnote{All proofs mentioned in this paper are machine-checked unless explicitly noted otherwise.} of basic laws: \hfill \lstinline{get $i$ empty = None}, \hfill
  \lstinline{get $i$ (set $i$ $v$ $m$) = Some $v$},\hfill \linebreak
  \lstinline{$i\not=j~\rightarrow$ get $i$ (set $j$ $v$ $m$) = get $i$ $m$}.
\item A purely functional implementation (which rules out hash tables, for example).
\item A \emph{persistent} semantics, so that \lstinline{set $j$ $v$ $m$} does not destroy $m$;  this arises naturally from the purely functional implementation.
\item Efficient representation of keys.
\item Efficient asymptotic time complexity of \lstinline{get} and \lstinline{set} operations, as extracted to OCaml.
\item Linear-time computation of the list of key-binding pairs of a finite-map, in sorted order by key, with proofs of its properties.
\item Linear-time combining two finite maps $m_1$ and $m_2$ with function $f$ yielding a map $m$
  such that, at any key $i$, $m(i)=f(m_1(i),m_2(i))$; with proofs of its properties.
\item[]~
\item[] \hspace*{-1em}\emph{Extended Requirements}
\item Efficiency in the face of sparseness:  that is, if the
  magnitude of individual keys is much greater than
  the number of keys bound in the map.
\item Efficiency not only as extracted to OCaml (i.e., with proofs erased),
  but also as represented in Coq and computed within Coq
  (i.e., without proof erasure).
\item Extensionality:  the property that $(\forall i.~m_1(i)=m_2(i))~\rightarrow~m_1=m_2$.  This allows Leibniz equality to be used in proofs about
  the larger systems in which the finite maps are embedded; otherwise,
  equivalence relations must be used, which is less convenient.
\end{enumerate}

The \emph{Core Requirements} were all satisfied by a
simple binary-trie data structure,
implemented in CompCert and used 2005--2021 \cite{CompCert-maps}.
In section \ref{core} we will present the data structure
and its operations, and discuss the proofs.

In more recent applications, the \emph{Extended Requirements} have
been needed.  Section~\ref{canonical} describes
a new variant of the data structure that has the extensionality
property, handles sparseness better, and
is significantly more efficient within Coq and somewhat more efficient
as extracted to OCaml.  It has been integrated in CompCert in 2021
\cite{CompCert-new-maps}.  
Source code for implementations and proofs may also be found in the
Coq development accompanying this article \cite{companion-dev}.

Extensionality could be achieved in other ways,
such as sigma-types in Coq (the dependent pair of a
data structure with a proof of its canonicity).  We describe
alternative implementations in section~\ref{other-approaches}.
Our benchmark measurements (section~\ref{sec:benchmarks})
show that our new variant, ``canonical binary tries''
performs well in all contexts, while each of the alternatives
is problematic in at least some contexts.

\paragraph{Related work.}
Many lookup-table data structures have been implemented and proved correct in
proof assistants.  The books \emph{Verified Functional Algorithms} \cite{appel17:vfa} and \emph{Functional Algorithms Verified} \cite{nipkow21:fav}
each describe several pure-functional tree data structures with proofs of
correctness (in Coq and Isabelle/HOL, respectively).
They describe binary search trees, red-black trees, 2-3 trees, AVL trees, Braun trees, binary tries, Patricia tries.
Of these,
\begin{itemize}
\item Only Braun trees, binary tries, and Patricia tries avoid the assumption
  that comparisons are constant-time.
\item Only red-black trees, 2-3 trees, AVL trees, and Patricia tries (but not Braun trees) handle 
  sparseness efficiently.
\item Only Braun trees have the extensionality property.  Braun trees
  support certain other operations, not supported by the
  other data structures (or by ours), useful for priority queues but
  not part of our core requirements.
\end{itemize}
That is, none of the standard data structures satisfy our Core and Extended Requirements. 

A different approach would be to build single-threaded mutable arrays
into the kernel of the logic, as was done long ago in ACL2 \cite{boyer02}
and more recently in Coq.  These do not satisfy \emph{persistence,}
except in a very inefficient way, nor sparseness.

\section{PTree: simple binary tries keyed by positive numbers}
\label{core}

A simple and efficient data structure for lookup tables (finite maps on
integer or string keys) is the \emph{hash table}.  But this is not suitable
for purely functional implementations, since the update operation
is destructive.\footnote{\emph{Persistent
arrays} give a functional semantics layered on an imperative
(destructive-update) implementation, but these are not suitable
because they would require extending the logic's kernel of axioms.}

Therefore one turns naturally to trees---for example,
binary search trees.  The lookup operator in a balanced BST
takes $\log N$ comparisons, and achieves asymptotic complexity
of $\log N$ per operation only if comparison takes constant time.
But Coq does not have a primitive integer type with constant-time
comparisons.\footnote{In fact, no programming language has
a primitive integer type with constant-time comparisons; they
typically have a \emph{range-limited} integer type $-2^k\le i < 2^k$
or a \emph{modular} integer type $(i \bmod 2^k)$.  
Most versions of Coq 2005-present have not had a 
built-in integer type with fast comparisons; and relying on
them would increase the size of the trusted base, that is,
Coq kernel code, Coq axioms, Coq extraction to OCaml, and OCaml
operations upon which we rely.}

Coq's logic, the Calculus of Inductive Constructions (CiC) encourages
the \emph{construction} of types and operations, instead of
\emph{axiomatization}.  For example, the natural numbers,
and addition upon them, are constructed as,
\begin{coq}
Inductive nat :=  O : nat | S : nat -> nat.
  
Fixpoint add ($n$ $m$ : nat) : nat := match $n$ with O => $m$ | S $p$ => S (add $p$ $m$) end.
\end{coq}
and then properties such as the associativity of \lstinline{add} are
proved as lemmas instead of being axiomatized.

But this representation of natural numbers is not suitable for use
as the keys of efficient lookup tables, because it is essentially
a \emph{unary} notation, in which representing the number $n$
takes space proportional to $n$, comparing $n<m$ takes time
proportional to $\mathrm{min}(n,m)$, and so on.  Therefore users of
Coq often use a \emph{binary} representation of integers,
constructed as follows:

\begin{coq}
Inductive positive := $~$ xI : positive -> positive $~$ | $~$ xO : positive -> positive $~$ | $~$ xH : positive.

Inductive Z := $~$  Z0 : Z $~$ | $~$ Zpos : positive -> Z $~$ | $~$ Zneg : positive -> Z.
\end{coq}

The ``meaning'' of \lstinline{xH} is the positive number 1;
\lstinline{xO($p$)} represents $2p+0$,
and \lstinline{xI($p$)} represents $2p+1$.
Therefore the number $13_{10}~=~1101_2~=~$\lstinline{xI(xO(xI xH))},
and the number 0 is unrepresentable---these numbers truly are
\emph{positive}, not merely nonnegative.

That means that the \lstinline{Z} type (of binary unbounded signed integers) 
has the property of \emph{extensionality}---no two data structures of
type \lstinline{Z} represent the same mathematical integer.  But the 
\lstinline{Z} type is of no further concern in this paper;
we will use \lstinline{positive} as the type of keys for lookup tables.

The number $p$, represented as a \lstinline{positive}, 
is represented with $\Theta(\log p)$ constructors.  When extracted
to OCaml, this really means $\log p$ two-word records (in which the first
word contains the constructor tag xI/xO/xH and the second word
is a pointer to the remainder).  That's a lot more than the single
word in which a native OCaml program or a C++ program might use
to represent an integer key, but it is \emph{efficient enough}
in practice, for the CompCert compiler and similar applications.

The positive number $p$, represented within Coq, also has $\log p$
constructors, but each Coq construction is represented within the Coq
kernel as two OCaml constructions that \emph{describe} the Coq
constructor.  So the number $p$ occupies $5 \log p$
words (or $8\cdot 5 \log p$ bytes on a 64-bit machine).
In practice this is usually efficient enough.\footnote{In OCaml,
an arity-$k$ constructor applied to arguments takes $k+1$ words,
including one for the constructor-tag.  In Coq, an arity-$k$ constructor
is represented as one arity-2 OCaml construction plus one length-$k$
array, which take (respectively) 3 words and $k+1$ words as represented
in OCaml within the Coq kernel.  The arity-2 OCaml construction
also points to a description of the constructor, but that description
is shared among all its uses so we don't count it.}

Given that Coq has this natural way to represent binary numbers using
inductive data types, it is natural to use \emph{binary tries} to
implement lookup tables.  A \emph{trie} is a directed tree in which every edge
is labeled from a finite alphabet; a \emph{word} (such as ``cat'') is looked
up by traversing the path of edges labeled sequentially with the letters
of the word (such as ``c'' then ``a'' then ``t''); and the \emph{nodes}
contain values of the range type of the lookup table.  A~\emph{binary}
trie is one in which the alphabet is $\{0,1\}$.

The implementation of binary tries, with \emph{positive} keys,
is straightforward in Coq:

\begin{coq}
Inductive option (A: Type) :=  Some : A -> option A | None : option A.

Inductive tree (A : Type) := Leaf : tree A | Node : tree A -> option A -> tree A -> tree A.

Definition empty {A : Type} : tree A := Leaf.

Fixpoint get {A : Type} ($i$ : positive) ($m$ : tree A) : option A :=
  match $m$ with
  | Leaf => None
  | Node $l$ $o$ $r$ =>  match $i$ with  xH => $o$ | xO $i'$ => get $i'$ $l$ | xI $i'$ => get $i'$ $r$ end
  end.

Fixpoint set {A : Type} ($i$ : positive) ($v$ : A) ($m$ : tree A) : tree A :=
  match $m$, $i$ with
  | Leaf, xH => Node Leaf (Some $v$) Leaf
  | Leaf, xO $i'$ => Node (set $i'$ $v$ Leaf) None Leaf
  | Leaf, xI $i'$ => Node Leaf None (set $i'$ $v$ Leaf)
  | Node $l$ $o$ $r$, xH => Node $l$ (Some $v$) $r$
  | Node $l$ $o$ $r$, xO $i'$ => Node (set $i'$ $v$ $l$) $o$ $r$
  | Node $l$ $o$ $r$, xI $i'$ => Node $l$ $o$ (set $i'$ $v$ $r$)
  end.
\end{coq}
That is, each node of the tree is either a \lstinline{Leaf},
meaning that \emph{no extension of the key leading to this node
is in the domain of the map},
or an internal \lstinline{Node} with left subtree $l$,
optional binding-at-this-node $o$,
and right subtree $r$.  If $o = \lst{None}$ then
the key leading to this node is not in the domain of the map,
but some extensions of it might be.  If $o = \lst{Some}~v$
then the key leading to this node is mapped to $v$,
and extensions of the key might also be mapped (in $l$ and $r$).
The edge from this node to subtree $l$ is implicitly labeled by
$0$ (or \lstinline{xO}), and the edge to subtree $r$ is
implicitly labeled by $1$ (or \lstinline{xI}).

Looking up a key $p$ in a trie takes time proportional to the
number of constructors representing the key, which is about
$\log_2 p$; update is similarly efficient; this is
\emph{efficient enough} for many important applications,
including CompCert.

From these definitions it is easy to prove the important
algebraic properties of maps:
\begin{coq}
Theorem gempty: (* ``get-empty'' *)
    forall (A: Type) ($i$: positive), get $i$ (empty A) = None.
Proof.  induction i; simpl; auto. Qed.

Theorem gss: (* ``get-set-same'' *)
  forall (A: Type) ($i$: positive) ($x$: A) ($m$: tree A), get $i$ (set $i$ $x$ $m$) = Some $x$.
Proof. induction i; destruct m; simpl; auto. Qed.

Theorem gso: (* ``get-set-other'' *)
    forall (A: Type) ($i$ $j$: positive) ($x$: A) ($m$: tree A), $i$ <> $j$ -> get $i$ (set $j$ $x$ $m$) = get $i$ $m$.
Proof.  induction i; intros; destruct j; destruct m; simpl;
        try rewrite <- (gleaf A i); auto; try apply IHi; congruence.
Qed.
\end{coq}
As usual in machine-checked proof scripts, the reader is not
necessarily expected to read and understand the proofs;
but the \emph{theorem statements} should be clear,
and \lstinline{induction i;} \lstinline{destruct m} suggests that the proof is by
induction on the positive number $i$ and then case analysis on $m$.

In less than 50 lines of Coq we have accomplished the first 6 of the core requirements: the operators \lstinline{empty, get, set,} proofs of the
algebraic laws, a purely functional implementation with a persistent
semantics, an efficient-enough representation of keys,
and efficient-enough asymptotic time complexity for many programs
extracted to OCaml.

Deletion (the \lstinline{remove} operation) is similar to insertion
(the \lstinline{set} operation) but runs into an issue: empty subtrees
can be represented in several different ways, and the most compact
representation (namely, the \lst{Leaf} constructor) should be favored.
\begin{coq}
Fixpoint remove {A: Type} ($i$: positive) ($m$: tree A) : tree A :=
  match $m$, $i$ with
  | Leaf, _ => Leaf
  | Node $l$ $o$ $r$, xH => Node' $l$ None $r$
  | Node $l$ $o$ $r$, xO $i'$ => Node' (remove $i'$ $l$) $o$ $r$
  | Node $l$ $o$ $r$, xI $i'$ => Node' $l$ $o$ (remove $i'$ $r$)
  end.
\end{coq}
In the second case, if $l$ and $r$ are both \lst{Leaf}, we
do not want to build a useless empty subtree \lst{Node Leaf None Leaf}.
This could also happen in the third and fourth cases if the recursive
call to \lst{remove} returns \lst{Leaf}.  Empty nonleaf subtrees are not
\emph{wrong,} but they waste memory.  So the solution is to use a
\lstinline{Node'} pseudoconstructor function, as follows:
\begin{coq}
Definition Node' {A: Type} ($l$: tree A) ($o$: option A) ($r$: tree A): tree A :=
  match $l$, $o$, $r$ with  Leaf, None, Leaf => Leaf  | _, _, _ => Node $l$ $x$ $r$ end.
\end{coq}
Crucially, the pseudoconstructor \lst{Node'} behaves like the actual
constructor \lst{Node} with respect to the \lst{get} operation:
\begin{coq}
Lemma gNode':
  forall (A: Type) ($i$: positive) ($l$: tree A) ($o$: option A) ($r$: tree A),
  get $i$ (Node' $l$ $o$ $r$) = match $i$ with xH => x | xO $i'$ => get $i'$ $l$ | xI $i'$ => get $i'$ $r$ end.
Proof.
  intros. destruct l, x, r; simpl; auto. destruct i; auto.
Qed.
\end{coq}
This makes it easy to prove \lst{get}-\lst{remove} algebraic properties
similar to the \lst{get}-\lst{set} properties.

To produce an association list of the key-binding pairs, one \emph{could}
write a function such as,
\begin{coq}
Fixpoint elements {A} ($m$: tree A) ($i$: positive) : list (positive * A) :=
 match $m$ with
 | Leaf => nil
 | Node $l$ $o$ $r$ => 
    elements $l$ (xO $i$)
   ++ match $o$ with None => nil | Some $x$ => (prev $i$, $x$)::nil  end
   ++ elements $r$ (xI $i$)
   end.
\end{coq}
where \lstinline{prev} is positive-reversal, 
so \lstinline{prev(xI (xI (xO (xH))) = xO (xI (xI xH))}.
However, this implementation is inefficient: since list-concatentation
\lstinline{++} takes time proportional to the length of its first
argument, this function will take about $N \log N$ time for a
tree containing $N$ elements.  So instead, one implements
\lstinline{elements} by accumulating the list to the right of the
current tree, costing time exactly linear in the number of nodes
in the tree.  We will not show the details.

Proofs of the properties of the \lstinline{elements} function are reasonably
straightforward, though more complex than the \lstinline{gso} theorem.

It is easy to define collective operations such as ``map'' and
``filter'' that transform a finite map into another finite map.  For
example, here is a \lstinline{map_filter $f$} operation that applies a
function \lstinline{$f$: $A$ $\rightarrow$ option $B$} to every value
$a$ contained in a \lstinline{tree $A$} map, discarding the value if
$f~a = \lst{None}$.

\begin{coq}
Fixpoint map_filter {A B} (f: A -> option B) (m: tree A) : tree B :=
  match m with
  | Leaf => Leaf
  | Node l None r => Node' (map_filter f l) None (map_filter f r)
  | Node l (Some a) r => Node' (map_filter f l) (f a) (map_filter f r)
  end.
\end{coq}
As in \lst{remove}, we use the pseudoconstructor \lst{Node'} to avoid generating
empty subtrees \lst{Node Leaf None Leaf}.

Many applications wish to combine two finite maps
\lstinline{$m_1:$ tree $A$, $m_2:$ tree $B$}
with a user-supplied function
\lstinline{$f$: option $A$ $\rightarrow$ option $B$ $\rightarrow$ option $C$}.
We assume that \linebreak \lstinline{$f$ None None = None},
and we require that \lstinline{combine $m_1$ $m_2$ = $m$}
if and only if \linebreak $\forall i, f(m_1(i))(m_2(i))=m(i)$.

\pagebreak[2]
One might propose an implementation something like this:
\begin{coq}
Fixpoint combine ($m_1$: tree A) ($m_2$: tree B) : tree C :=
 match $m_1$, $m_2$ with
 | Leaf, Leaf => Leaf
 | Leaf, Node $l_2$ $o_2$ $r_2$ => Node' (combine Leaf $l_2$) (f None $o_2$) (combine Leaf $r_2$)
 | Node $l_1$ $o_1$ $r_1$, Leaf => Node' (combine $l_1$ Leaf) (f $o_1$ None) (combine $r_1$ Leaf)
 | Node $l_1$ $o_1$ $r_1$, Node' $l_2$ $o_2$ $r_2$ => Node (combine $l_1$ $l_2$) (f o1 $o_2$) (combine $r_1$ $r_2$)
 end.
\end{coq}
but there is a problem here.  Some of the recursive
calls use substructures of \lstinline{$m_1$}
while others use substructures of \lstinline{$m_2$},
so this is not actually legal in Coq.  The solution
is that once a \lstinline{Leaf} is reached on either side, we can
traverse the other side using \lst{map_filter}, which is recursive on just
one tree.

\begin{coq}
Fixpoint combine ($m_1$: tree A) ($m_2$: tree B) : tree C :=
  match $m_1$, $m_2$ with
  | Leaf, _ => map_filter (fun b => f None (Some b)) $m_2$
  | _, Leaf => map_filter (fun a => f (Some a) None) $m_1$
  | Node $l_1$ $o_1$ $r_1$, Node $l_2$ $o_2$ $r_2$ => Node' (combine $l_1$ $l_2$) (f $o_1$ $o_2$) (combine $r_1$ $r_2$)
end.
\end{coq}

This \lstinline{combine} function is quite general and reasonably
efficient.  It takes time linear in the total sizes of the
input trees.  Proofs of its properties are straightforward,
for example,

\begin{coq}
Theorem gcombine: forall ($m_1$: tree A) ($m_2$: tree B) ($i$: positive),
      get $i$ (combine $m_1$ $m_2$) = f (get $i$ $m_1$) (get $i$ $m_2$).
\end{coq}
The proof is only 6 lines of Coq.

\paragraph{Summary of Section~\ref{core}.}
Binary tries, with keys that are constructive binary positive numbers,
are a simple data structure with an easy implementation and reasonably
straightforward proofs.  They are efficient enough in practice
for many useful applications, especially as extracted into OCaml.

\section{The need for extensionality}
\label{sec:need}

\emph{Extensionality} is a property of equality in formal logics.  It
says that two objects are equal if they have the same external
properties.  For example, the extensionality property for functions
says that two functions $f$ and $g$ are equal as soon as they have the
same domain and satisfy $\forall x,~f(x) = g(x)$.  This extensionality
property holds in set theory but not in Coq's type theory.

For finite maps, the extensionality property says that two maps are
equal if they are equivalent, that is, if they map equal keys to equal
values.  Like most popular implementations of finite maps, binary
tries (represented as in Section~\ref{core}) are not extensional.
That is, equivalent tries are not necessarily equal, as shown by the
following example.

\[
\begin{array}{r@{\qquad}r@{\qquad\qquad}l}
m_1& \cdot~1~((\cdot~2~\cdot)\,\circ\,(\cdot~2~\cdot))
   &\cdot~\mathrm{represents}~\mathsf{Leaf}
\\
m_2& (\cdot\,\circ\,\cdot)~1~((\cdot~2~\cdot)\,\circ\,(\cdot~2~\cdot))  
   & \circ~\mathrm{represents}~\mathsf{None}
\end{array}
\]
Here, $m_1$ and $m_2$ are equivalent (for any $i$,
\lstinline{get $i$ $m_1$ = get $i$ $m_2$}),
but $m_2$ has an extra empty left subtree 
\lstinline{Node Leaf None Leaf} where $m_1$ has just \lstinline{Leaf}.

This is not an insurmountable problem.  One can reason using
equivalence rather than equality.
Coq has support
for such ``Setoid'' reasoning;  but it's still inconvenient,
and a truly canonical representation of tries would
simplify many definitions and proofs.  For example, consider the
following two algebraic laws of finite maps:
\begin{eqnarray*}
\lst{set}~k~v~m & = & m \qquad\mbox{if $\lst{get}~k~m = \lst{Some}~v$}
\\
\lst{set}~k_1~v_1~(\lst{set}~k_2~v_2~m) & = & \lst{set}~k_2~v_2~(\lst{set}~k_1~v_1~m)
\qquad\mbox{if $k_1 \not= k_2$}
\end{eqnarray*}
If the representation of $m$ were canonical
(which the binary tries of Section~\ref{core} are not),
then these laws would be trivial consequences of extensionality.
In the absence of extensionality,
these laws require specific proofs, using the
definition of \lst{set} and induction over $m$.

More generally, a large Coq development might have types that contain finite maps, such as a type for ``program modules'' in which finite maps represent bindings of names to definitions.  If the finite maps do not support
Leibnitz equality, and one must therefore use setoids, then
equality reasoning about 
``program modules'' must also use setoids.

Coq proofs 
with setoids are larger and slower than proofs with Leibnitz equality.
Extensional tries (permitting Leibnitz equality) could perform much
faster in Coq proofs.  And indeed they do; see section~\ref{sec:impactVST}.

One might think, ``extra empty subtrees will never arise in practice,
since the \lst{set} operation cannot produce them.''  But
then one would have to maintain the invariant that a given trie
was produced by a sequence of \lst{set} operations.

One might think, ``we'll maintain that invariant using dependent types
containing proofs that the tree has no empty subtrees.''
Section~\ref{other-approaches} describes two attempts along this
line and the efficiency problems they cause.

Therefore, what we want is a \emph{first-order, naturally
extensional representation} of tries.

\section{Canonical binary tries}
\label{canonical}

For a first-order naturally extensional representation that
handles sparseness better, 
our solution is to make an inductive datatype that \emph{cannot}
represent an empty mapping.  The data structure is as follows:

\begin{coq}
Inductive tree' (A: Type) : Type := 
 | Node001: tree' A -> tree' A  $\hspace{3cm}$(* only a right subtree *)
 | Node010: A -> tree' A        $\hspace{3.65cm}$(* only a middle value *)
 | Node011: A -> tree' A -> tree' A  $\hspace{2.3cm}$(* only middle and right *)
 | Node100: tree' A -> tree' A        $\hspace{3cm}$(* only a left subtree *)
 | Node101: tree' A -> tree' A -> tree' A   $\hspace{1.55cm}$ (* left, right, no middle *)
 | Node110: tree' A -> A -> tree' A   $\hspace{2.2cm}$   (* only left and middle *)
 | Node111: tree' A -> A -> tree' A -> tree' A.  $\hspace{0.85cm}$(* left, middle, right *)

Inductive tree (A: Type) : Type := 
  | Empty : tree A
  | Nodes: tree' A -> tree A.
\end{coq} 
Each of the three digits in the name of a node constructor
represents the presence or absence of the
left, middle, or right components of the node.
Thus, \lstinline{Node101 $l$ $r$} represents a node
with left and right subtrees but no middle,
and \lstinline{Node011 $x$ $r$} represents
what would have been written (in our previous representation)
as
\lstinline{Node Leaf (Some $x$) $r$}.

A crucial property of this data structure is \emph{canonicity}:
for any finite set $S$ of (key, value) bindings,
there is only one value of type \lstinline{tree} that
represents this set.  If $S$ is empty, it is represented by \lstinline{Empty},
and cannot be represented  by \lstinline{Nodes $t$}, as a value $t$ of
type \lstinline{tree'} always contains at least one (key, value) binding.
If $S$ is not empty, it can only be represented by
\lstinline{Nodes $t$}.  The head constructor of $t$ is uniquely
determined by whether the sets
$$
S_h = \{ (\lst{xH}, v) \mid (\lst{xH}, v) \in S \} \quad
S_o = \{ (i, v) \mid (\lst{xO}~i, v) \in S \} \quad
S_i = \{ (i, v) \mid (\lst{xI}~i, v) \in S \}
$$
are empty or not.  Induction on the size of $S$ shows that the
subtrees of $t$, if any, are uniquely determined.

Canonicity implies extensionality, of course: two values of type
\lstinline{tree} that represent the same set of (key, value) pairs are
equal.  In addition, canonicity improves computational efficiency
compared with the noncanonical binary tries from section~\ref{core}:
no memory is consumed to represent empty left, middle, or right
components of a node.  As the experimental evaluation in
sections~\ref{sec:benchmarks}, \ref{sec:CompCert} and~\ref{sec:impactVST} shows,
the gains in memory usage and execution times are significant,
especially for sparse maps.

Defining the lookup and update functions is not quite as simple as before:

\begin{coq}
Definition empty (A : Type) := (Empty : tree A).

Fixpoint get' {A} ($p$: positive) ($m$: tree' A) : option A :=
  match $p$, $m$ with
  | xH, Node001 _ => None
  | xH, Node010 $x$ => Some $x$
  | xH, Node011 $x$ _ => Some $x$
  | xH, Node100 _ => None
  | xH, Node101 _ _ => None
  | xH, Node110 _ $x$ => Some $x$
  | xH, Node111 _ $x$ _ => Some $x$
  | xO $q$, Node001 _ => None
  | xO $q$, Node010 _ => None
  | xO $q$, Node011 _ _ => None
  | xO $q$, Node100 $m'$ => get' $q$ $m'$
  | xO $q$, Node101 $m'$ _ => get' $q$ $m'$
  | xO $q$, Node110 $m'$ _ => get' $q$ $m'$
  | xO $q$, Node111 $m'$ _ _ => get' $q$ $m'$
  | xI $q$, Node001 $m'$ => get' $q$ $m'$
  | xI $q$, Node010 _ => None
  | xI $q$, Node011 _ $m'$ => get' $q$ $m'$
  | xI $q$, Node100 $m'$ => None
  | xI $q$, Node101 _ $m'$ => get' $q$ $m'$
  | xI $q$, Node110 _ _ => None
  | xI $q$, Node111 _ _ $m'$ => get' $q$ $m'$
end.
 
Definition get {A} ($p$: positive) ($m$: tree A) : option A :=
  match $m$ with Empty => None | Nodes $m'$ => get' $p$ $m'$ end.
\end{coq}
The \lstinline{get'} function has 21 cases,
handling each combination of one of the 3 \lstinline{positive}
constructors and one of the 7 node constructors.

The update function \lst{set} follows the same pattern.  An update to the
empty tree is handled by the \lst{set0} function, which creates a nonempty
tree with a single branch:

\begin{coq}
  Fixpoint set0 {A} (p: positive) (x: A) : tree' A :=
  match p with
  | xH => Node010 x
  | xO q => Node100 (set0 q x)
  | xI q => Node001 (set0 q x)
  end.
\end{coq}
To update a nonempty tree, the function \lst{set'} performs the same 21-case
analysis as \lst{get'}:
\begin{coq}
  Fixpoint set' {A} (p: positive) (x: A) (m: tree' A) : tree' A :=
  match p, m with
  | xH, Node001 r => Node011 x r
  | xH, Node010 _ => Node010 x
  | xH, Node011 _ r => Node011 x r
  | xH, Node100 l => Node110 l x
  | xH, Node101 l r => Node111 l x r
  | xH, Node110 l _ => Node110 l x
  | xH, Node111 l _ r => Node111 l x r
  | xO q, Node001 r => Node101 (set0 q x) r
  | xO q, Node010 y => Node110 (set0 q x) y
  | xO q, Node011 y r => Node111 (set0 q x) y r
  | xO q, Node100 l => Node100 (set' q x l)
  | xO q, Node101 l r => Node101 (set' q x l) r
  | xO q, Node110 l y => Node110 (set' q x l) y
  | xO q, Node111 l y r => Node111 (set' q x l) y r
  | xI q, Node001 r => Node001 (set' q x r)
  | xI q, Node010 y => Node011 y (set0 q x)
  | xI q, Node011 y r => Node011 y (set' q x r)
  | xI q, Node100 l => Node101 l (set0 q x)
  | xI q, Node101 l r => Node101 l (set' q x r)
  | xI q, Node110 l y => Node111 l y (set0 q x)
  | xI q, Node111 l y r => Node111 l y (set' q x r)
  end.

  Definition set {A} (p: positive) (x: A) (m: tree A) : tree A :=
  match m with
  | Empty => Nodes (set0 p x)
  | Nodes m' => Nodes (set' p x m')
  end.
\end{coq}

Proofs of the fundamental theorems (such as \lstinline{gss}
and \lstinline{gso}) are still short and easy:
about 25 lines in all, including supporting lemmas.

Other functions that operate over a single tree remain reasonably easy
to write.  For example, here is the recursive part of the \lst{map_filter}
function that applies a function \lstinline{$f$: $A$ $\rightarrow$ option $B$}
to a nonempty tree \lstinline{$m$: tree' $A$}:
\begin{coq}
  Fixpoint map_filter' {A B} (f: A -> option B) (m: tree' A) : tree B :=
    match m with
    | Node001 r => Node Empty None (map_filter' f r)
    | Node010 x => Node Empty (f x) Empty
    | Node011 x r => Node Empty (f x) (map_filter' f r)
    | Node100 l => Node (map_filter' f l) None Empty
    | Node101 l r => Node (map_filter' f l) None (map_filter' f r)
    | Node110 l x => Node (map_filter' f l) (f x) Empty
    | Node111 l x r => Node (map_filter' f l) (f x) (map_filter' f r)
    end.
\end{coq}

The \lst{Node} function used in the definition above is similar to the
\lst{Node} constructor and the \lst{Node'} pseudoconstructor of
section~\ref{core}: given a possibly empty left subtree, a possibly
absent value, and a possibly empty right subtree, it returns the
corresponding tree.

\begin{coq}
  Definition Node {A} (l: tree A) (o: option A) (r: tree A) : tree A :=
    match l,o,r with
    | Empty, None, Empty => Empty
    | Empty, None, Nodes r' => Nodes (Node001 r')
    | Empty, Some x, Empty => Nodes (Node010 x)
    | Empty, Some x, Nodes r' => Nodes (Node011 x r')
    | Nodes l', None, Empty => Nodes (Node100 l')
    | Nodes l', None, Nodes r' => Nodes (Node101 l' r')
    | Nodes l', Some x, Empty => Nodes (Node110 l' x)
    | Nodes l', Some x, Nodes r' => Nodes (Node111 l' x r')
    end.
\end{coq}

The case analysis performed by \lst{Node} can be partially redundant.  For
instance, in the first case of the \lst{map_filter'} function, we use \lst{Node}
with \lst{l = Empty} and \lst{o = None}; we would prefer to avoid
discriminating over \lst{l} and \lst{o} in \lst{Node}.  This is easily achieved by
asking Coq to unfold the definition of \lst{Node} inside \lst{map_filter'}, then
simplify using call-by-value evaluation:

\begin{coq}
  Definition fast_map_filter' := Eval cbv [Node] in @map_filter'.
\end{coq}

The resulting function performs the minimal number of case
distinctions on the results of \lst{f} and of recursive calls.  Hence, it
executes faster than \lst{map_filter'}, both within Coq and after extraction
to OCaml.  At the same time, \lst{fast_map_filter'} is \emph{convertible}
with \lst{map_filter'}, since it is obtained by reductions, hence the two
functions are definitionally equal, and all the results proved on the
nicer-looking \lst{map_filter'} hold for the faster \lst{fast_map_filter'}.

Functions that operate over two trees, such as the \lst{combine} function
from Section~\ref{core}, are more difficult to write because of the
high number of cases.  Consider the recursive case of the \lst{combine}
function, the one that operates on two nonempty trees in parallel.

\begin{coq}
  Variables A B C: Type.
  Variable f: option A -> option B -> option C.
  Definition combine'_l := map_filter' (fun a => f (Some a) None).
  Definition combine'_r := map_filter' (fun b => f None (Some b)).

  Fixpoint combine' (m1: tree' A) (m2: tree B) {struct m1} : tree C :=
    match m1, m2 with
    | ...
    | Node101 l1 r1, Node011 x2 r2 => Node (combine'_l l1) (f None (Some x2)) (combine' r1 r2)
    | ...
    end.
\end{coq}
A naive case analysis on \lst{m1} and \lst{m2} results in 49 cases.
Each case is simple enough: we call \lst{combine'} if there
are two left subtrees or two right subtrees, \lst{combine'_l} if there is
only a left subtree, \lst{combine'_r} if there is only a right subtree.
Likewise, reasoning about these 49 cases is relatively easy, using
Coq's tactic language for automation.  But writing all the cases in
the first place is a pain!

Even so, \emph{proofs} about the 49-case \lstinline{combine'} function
can be concise.  Consider this theorem, about the interaction
of \lst{combine'} with \lst{get'}:
\begin{coq}
Lemma gcombine': forall m1 m2 i,
        get i (combine' m1 m2) = f (get' i m1) (get' i m2).
Proof.
    induction m1; destruct m2; intros; simpl; rewrite gNode;
    destruct i; simpl; auto using gcombine'_l, gcombine'_r.
Qed.
\end{coq}
Tactics make it easy to chain
induction on \lst{m1}, case analysis (destruct) on
\lst{m2}, simplification, rewriting, and case analysis on \lst{i}, all in
two lines of Ltac scripting.

\section{Defining functions with case analysis, in Coq}

Our 49-case \lstinline{combine'} function has a regular structure
that is motivated by the way that canonical tries relate to
original tries.  
We briefly digress to show how to define
such functions---that do a large (but regularly structured) case
analysis---concisely in Coq.
We show two approaches:
tactical function definition, and \emph{views}.

\subsection{Tactical function definition.}
Here is a definition \lstinline{combine'_by_tac} that
produces the same function as the 49-case \lstinline{combine'}
described above:
\begin{coq}
Fixpoint combine'_by_tac (m1: tree' A) (m2: tree' B) {struct m1} : tree C.
Proof.
  destruct m1 as [ r1 | x1 | x1 r1 | l1 | l1 r1 | l1 x1 | l1 x1 r1 ];
  destruct m2 as [ r2 | x2 | x2 r2 | l2 | l2 r2 | l2 x2 | l2 x2 r2 ];
  (apply Node;
   [ solve [ exact (combine'_by_tac l1 l2)
           | exact (combine'_l l1)
           | exact (combine'_r l2)
           | exact Empty]
   | solve [ exact (f (Some x1) (Some x2))
           | exact (f None (Some x2))
           | exact (f (Some x1) None)
           | exact None]
   | solve [ exact (combine'_by_tac r1 r2)
           | exact (combine'_l r1)
           | exact (combine'_r r2)
           | exact Empty ]
   ]).
Defined.
\end{coq}
This definition is in the same context that binds parameters
\lst{A,B,C,f} as the definition above of \lst{combine'}.
The first line describes a \emph{goal:} define a recursive function
of a particular type by structural induction on \lst{m1}.
The ``Proof'' is a case analysis described in Coq's tactic language:
the \lstinline{destruct} builds a pattern-\lstinline{match}
into the ``proof'' (in this case,
into the ``program'').  So we start with two nested 7-case matches,
that is, a 49-case double pattern match.  Each one of those
has a right-hand-side that's an application of the \lstinline{Node}
pseudo constructor, so we say \lstinline{apply Node} in
the tactic language.   Since \lstinline{Node} is a 3-argument
function, we have three subgoals remaining,
separated by the punctuation \lstinline{[$~~$|$~~$|$~~$]}.
In each case, we try four different things (the \lstinline{exact} tactics),
in the order given.  The first of the four that type-checks is
the one that will be chosen for use in defining
the function.

The reader may not be confident that this defines the right function!
Indeed, the author of the tactic may not be confident.  One can gain
confidence by this theorem, which shows that the new implementation
is $\beta\!\delta$-convertible with the hand-written 49-case version:
\begin{coq}
Lemma combine'_by_tac_eq: combine'_by_tac = combine'.
Proof. reflexivity. Qed.
\end{coq}

However, the purpose of this tactic is to \emph{avoid} having to
write the 49-case version by hand.  So a more useful theorem
to prove correctness of the tactical implementation (or any
implementation) is that \lstinline{combine'} interacts
properly with \lstinline{get'}.  And indeed, the
same \lst{gcombine'} lemma-statement (as shown
in \S\ref{canonical})
\emph{and exactly the same
tactical proof} works to prove that theorem
for \lst{combine'_by_tac} as for the direct implementation
of \lst{combine'}.

\subsection{Views}
Instead of definition-by-tactic,
another way to avoid explosion in the number of cases is to work
with canonical binary tries using a \emph{view} consisting of two
cases only: either an empty leaf or a node consisting of a left
subtree, an optional value, and a right subtree.  In other words, the
view is isomorphic to the concrete representation of the original
binary tries from section~\ref{core}.  One half of this view---the
one that lets us construct values of type \lst{tree A}---consists of
the \lst{Empty} constructor and the \lst{Node} function mentioned above:
\begin{coq}
    Empty : forall A, tree A
    Node  : forall A, tree A -> option A -> tree A -> option A
\end{coq}
The other half of the view---the one that lets us inspect and
recurse over values of type \lst{tree A}---is provided by elimination
and induction principles.  Here is a simple, nonrecursive case analysis with
two cases:
\begin{coq}
Definition tree_case {A B} (empty: B)
                           (node: tree A -> option A -> tree A -> B)
                           (m: tree A) : B :=
  match m with
    | Empty => empty
    | Nodes (Node001 r) => node Empty None (Nodes r)
    | Nodes (Node010 x) => node Empty (Some x) Empty
    | Nodes (Node011 x r) => node Empty (Some x) (Nodes r)
    | Nodes (Node100 l) => node (Nodes l) None Empty
    | Nodes (Node101 l r) => node (Nodes l) None (Nodes r)
    | Nodes (Node110 l x) => node (Nodes l) (Some x) Empty
    | Nodes (Node111 l x r) => node (Nodes l) (Some x) (Nodes r)
    end.
\end{coq}
We expect this case analysis principle to satisfy the two equations
\begin{coq}
    tree_case empty node Empty = empty
    tree_case empty node (Node l o r) = node l o r
\end{coq}
The second equation is not true in general: in the case where
\lst{l} and \lst{r} are \lst{Empty} and \lst{o} is \lst{None}, \lst{Node l o r} is \lst{Empty} and
the \lst{tree_case} analysis returns \lst{empty},
whereas our second equation requires it to return \lst{node Empty None Empty}.
To support the
equation, we can require that \lst{node Empty None Empty = empty}, showing
that the two cases of the analysis agree in this special case.
Alternatively, we can rule out the special case by adding the
precondition \lst{not_trivially_empty l o r} to the second equation.
The \lst{not_trivially_empty} predicate is defined as
\begin{coq}
Definition not_trivially_empty {A} (l: tree A) (o: option A) (r: tree A) :=
  match l, o, r with
  | Empty, None, Empty => False
  | _, _, _ => True
  end.
\end{coq}
The custom induction principle that we define later in this section
provides us with the \lst{not_trivially_empty l o r} guarantee in the
\lst{Node l o r} case.  Hence we can always apply the \lst{tree_case}
equations when reasoning with this induction principle.

A minor variant of \lst{tree_case} implements a recursion principle.  We
still have two cases, \lst{empty} and \lst{node}, but the \lst{node} case also
receives the results of recursing over the left and right subtrees.
\begin{coq}
  tree_rec: forall {A B} (empty: B)
                   (node: tree A -> B -> option A -> tree A -> B -> B),
                   tree A -> B.
\end{coq}

Along the same lines, we can also define a general, dependently typed
induction principle, usable both for computations and for proofs.
Its type is similar to that of the Coq-generated induction
principle for the two-constructor \lst{tree} type of section~\ref{core},
except that the \lst{node} case also receives a proof of the
\lst{not_trivially_empty} guarantee:
\begin{coq}
  tree_ind: forall {A: Type} (B: A -> Type)
                   (empty: B Empty)
                   (node: forall l, B l -> forall o r, B r -> not_trivially_empty l o r -> B (Node l o r)),
                   (m: tree A), B m.
\end{coq}

One use for this induction principle is to simplify proofs about
hand-written functions over type \lst{tree}.  For example, to prove
properties of the \lst{set} function defined above, we can first show the
following equations:
\begin{coq}
          set xH v Empty = Node Empty (Some v) Empty
      set (xO q) v Empty = Node (set q v Empty) None Empty
       set (xI q) v Empty = Node Empty None (set q v Empty)
     set xH v (Node l o r) = Node l (Some v) r
 set (xO q) v (Node l o r) = Node (set q v l) o r
  set (xI q) v (Node l o r) = Node l o (set q v r)
\end{coq}
which follow by a trivial case analysis on \lst{l}, \lst{o} and \lst{r}.
Then, we can prove properties such as
$\lst{get}~i~(\lst{set}~i~v~m) = \lst{Some}~v$
by an outer structural induction over $i$ and an inner induction over
$m$ that uses the \lst{tree_ind} custom induction principle.  This reduces
the number of cases to consider to $3 \times 2 = 6$ cases.

Custom case analysis and induction principles can also be used to
define functions over trees.  For instance, the infamous \lst{combine}
function can be defined quite concisely by:
\begin{coq}
  tree_rec combine_r
           (fun l1 rec_l1 o1 r1 rec_r1 m2 =>
             tree_case
               (combine_l (Node l1 o1 r1))
               (fun l2 o2 r2 => Node (rec_l1 l2) (f o1 o2) (rec_r1 r2))
               m2)
\end{coq}

This definition is much more compact than
the hand-written, 49-case definition.  However, the \lst{tree_rec}
and \lst{tree_case} functions add significant run-time overhead.  We can
recover most of the efficiency of the hand-written version by partial
evaluation within Coq: as shown above for \lst{map_filter'}, we can use
\lst{Eval cbv} to force Coq to expand the \lst{tree_rec} and \lst{tree_case}
functions, then simplify.  Even then, a minor inefficiency remains:
the second argument to \lst{combine} is tested for emptiness at each
recursive step over the first argument, instead of being tested once
and for all in the hand-written or tactic-based implementations.  A
custom double induction principle, operating over two trees at once,
can remove this last inefficiency.

To conclude this discussion: the canonical representation of binary
tries greatly increases the number of cases in function
definitions and in proofs.  This increase can be challenging for
functions operating on two or more tries at the same time, but can be
managed either by writing tactics that define these functions, or by
defining appropriate case analysis and induction principles that
reduce the number of cases, then using Coq's built-in evaluation
facilities to remove the execution overhead caused by these principles.
In our experience, the former approach based on views make it easy to
define correct functions on the first try, while the tactics-based
approach requires more trial and error.  However, tactics are able to produce
function definitions that have exactly the required shape for optimal
performance; this is not always the case with partial evaluation of
definitions that use views, which provides less control on the exact
shape of the final definition.

\section{Other approaches to extensionality}
\label{other-approaches}

The canonical, first-order representation we have just presented in
section~\ref{canonical} is not the only way to achieve
extensionality: another way is to use dependent types, as we now
demonstrate.
The reader may choose to skip this section, because it turns out that
none of these other approaches performs as well, in a wide enough range of
applications, as the canonical representation.

\subsection{Subset types}
\label{subset-types}

The simple
tries from section~\ref{core} are not extensional
because they might contain
subtrees \lst{Node Leaf None Leaf}, which contain no elements
but structurally differ from \lst{Leaf}.  The extensionality
property holds if we restrict ourselves to trees that satisfy the
following invariant \lst{wf}, ``well-formed: the tree contains no empty nodes
\lst{Node Leaf None Leaf}.''
\begin{coq}
Definition not_trivially_empty {A} (l: tree A) (o: option A) (r: tree A) :=
  match l, o, r with
  | Leaf, None, Leaf => False
  | _, _, _ => True
  end.

Inductive wf {A}: tree A -> Prop :=
  | wf_Leaf: wf Leaf
  | wf_Node: forall l o r,  wf l -> wf r -> not_trivially_empty l o r -> wf (Node l o r).
\end{coq}
It is not difficult to show that if \lstinline{wf $m_1$} and
\lstinline{wf $m_2$} hold, and if $m_1$ and $m_2$ are equivalent (for any $i$,
\lstinline{get $i$ $m_1$ = get $i$ $m_2$}), then $m_1$ and $m_2$ are
structurally equal.

To maintain the invariant through all operations over trees, we define
the type of tries $m$ that satisfy \lstinline{wf $m$} as a Coq subset
type, also called sigma-type:
\begin{coq}
Definition t (A: Type) : Type := { m: tree A | wf m }.
\end{coq}
Values of this type are dependent pairs of a tree $m$ and a proof of
\lstinline{wf $m$}.  This approach is used in the Std++ library \cite{Stdpp-pmap}.
Earlier, it was used by Filliâtre and Letouzey in their
implementations of finite sets and finite maps by AVL trees
\cite{DBLP:conf/esop/FilliatreL04},
to maintain the invariant that all trees considered are binary search
trees.

A strength of this approach is that it reuses the previous
implementation of tries (definitions of operations and proofs of properties),
simply wrapping them up with proofs of the \lst{wf} invariant.
\begin{coq}
Fixpoint set_raw {A : Type} ($i$ : positive) ($v$ : A) ($m$ : tree A) : tree A :=
  match m with $\ldots$ end.

Lemma set_wf: forall {A: Type} i (v: A) {m: tree A}, wf m -> wf (set_raw i v m).
Proof. $\ldots$ Qed.

Definition set {A: Type} (i: positive) (v: A) (m: t A) : t A :=
  exist _ (set_raw i v (proj1_sig m))  (set_wf i v (proj2_sig m)).
\end{coq}

The ``subset type'' approach executes efficiently after extraction:
propositions and proofs are erased during extraction, hence the type
\lst{t} is identical to \lst{tree} in the extracted
OCaml code, and there is no overhead compared with the original
implementation of tries.

However, execution within Coq is inefficient because the proof of
\lstinline{wf $m$} grows linearly \emph{in the number of operations
  performed over $m$}.  In effect, the proof part of a value of type
\lstinline{t A} contains the history of all operations applied
to produce this value.

Consider for instance the result of setting the key \lst{xH}
to values $v_1, \ldots, v_n$ successively.  The value part is the
small tree \lstinline{Node Leaf (Some $v_n$) Leaf}, but the proof part
is\linebreak \lstinline{set_wf xH $v_n$ ($\ldots$ (set_wf xH $v_1$ wf_Leaf) $\ldots$)},
a term of size $O(n)$.  Assuming the proof of lemma
\lst{set_wf} is opaque, this proof term does not reduce and
takes $O(n)$ space and $O(n)$ time when normalization is requested.

We experimented with alternate, ``transparent'' definitions of
\lst{set_wf} and related lemmas that would reduce to small proof
terms during normalization, but failed to find definitions that
would be efficient both in space and in normalization time.

In logics (such as HOL and Isabelle/HOL)
that use the LCF approach \cite{gordon79}
to the representation of proofs---i.e., that do not represent
proof terms at all---these ``subset types'' are likely to be
quite efficient. But Coq represents proof trees explicitly. 
If only the proof terms of these sigma types could be erased
in computation within Coq, as they are erased in extraction,
then the subset-type implementation might perform very well inside Coq.
An early implementation of Agda followed this approach of erasing
irrelevant terms in the internal syntax, following the model-theoretic
study by Abel, Coquand and Pagano \cite{Abel11}.  It was
later found to cause problems with higher-order unification and 
abandoned.\footnote{Andreas Abel, personal communications, Aug-Sept 2022.
Examples of problematic unifications can be found at \url{https://github.com/agda/agda/issues/483}}
More recently, Pujet and Tabareau \cite{pujet22} have worked out
another type theory of this kind of organic computational
proof-irrelevance.
If it were implemented in the Coq kernel, then many kinds of data
structures could be made extensional using sigma types.

\subsection{A poor man's inductive-inductive definition}
\label{GADT}

Rather than maintaining the \lst{wf} invariant as a separate proof term, as
in section~\ref{subset-types}, we could try to make it part of the
main tree data structure, along the following lines:
\begin{coq}
Inductive tree (A : Type) :=
  | Leaf : tree A
  | Node : forall (l: tree A) (o: option A) (r: tree A), $~$ not_trivially_empty l o r -> tree A.
\end{coq}
The \lst{Node} constructor carries not only an optional value
and left and right subtrees, but also a proof that the node is not
trivially empty.  We would expect this representation to compute
within Coq more efficiently than the subset type representation, since
proofs of \lst{not_trivially_empty} are small, and only those proofs
relevant to the nodes of the tree are kept.

The definition above is incorrect, of course, since the
\lst{not_trivially_empty} predicate mentions the \lst{Leaf}
constructor of the type \lst{tree} that we are defining.  We
need to define the predicate and the tree simultaneously, along the
following lines:
\begin{coq}
Inductive tree (A : Type) :=
  | Leaf : tree A
  | Node : forall (l: tree A) (o: option A) (r: tree A), $~$ not_trivially_empty l o r -> tree A.

with not_trivially_empty {A: Type}: tree A -> option A -> tree A -> Prop :=
  | not_trivially_empty_intro: forall l o r,
           ~(l = Leaf /\ o = None /\ r = Leaf) -> not_trivially_empty l o r.
\end{coq}
This is called an \emph{inductive-inductive definition}: the type
\lst{tree} is defined mutually recursively with the type family
\lst{not_trivially_empty} that is indexed over type \lst{tree}.

Such inductive-inductive types are supported in Agda but not in Coq.
However, our use of induction-induction is so simple that we can work
around this limitation of Coq.  We index the type of trees with an
additional parameter of type \lst{kind}, with the kind \lst{Empty} standing
for the constructor \lst{Leaf}, and the kind \lst{Nonempty} standing for the
constructor \lst{Node}:
\begin{coq}
Inductive kind : Type := Empty | Nonempty.

Definition not_trivially_empty {A: Type} (kl: kind) (o: option A) (kr: kind) :=
  ~(kl = Empty /\ o = None /\ kr = Empty).

Inductive tree (A : Type) : kind -> Type :=
  | Leaf : tree A Empty
  | Node : forall (kl kr: kind) (l: tree A kl) (o: option A) (r: tree A kr),
                  not_trivially_empty kl o kr -> tree A Nonempty.
\end{coq}
This breaks the circularity in the definition of \lst{not_trivially_empty} and
turns \lst{tree} into a regular inductive type.  Finally, a well-formed
tree is defined as a dependent pair of a kind and a tree of that kind:
\begin{coq}
Inductive t (A: Type) : Type :=
  | Tree : forall (k: kind), tree A k -> t A.
\end{coq}
With some elbow grease, we can adapt the definitions of operations
over trees, first at type \lst{tree A k}, then at type \lst{t A} after
abstracting over the kind \lst{k}.  For example, the \lst{set} operation looks
as follows:
\begin{coq}
Fixpoint set' {A: Type} (k: kind) (i: positive) (v: A) (m: tree A k) : tree A Nonempty := ...

Definition set {A: Type} (i: positive) (v: A) (m: t A) : t A :=
  let '(Tree k m) := m in Tree Nonempty (set' k i v m).
\end{coq}

Concerning efficiency of evaluation, 
canonical binary tries seem to have the same asymptotic time and space
complexity as the original, nonextensional implementation from
section~\ref{core}, both after extraction and within Coq---see Table~\ref{t:benchs}.
However,
the \lst{Node} constructor is bigger by a constant factor: in Coq, it
additionally carries two kinds (\lst{kl} and \lst{kr}) and a proof of
\lst{not_trivially_empty}; after extraction to OCaml, it still carries the two
kinds \lst{kl} and \lst{kr}, even if they do not participate in the
computations.  Consequently, the constant factors for space and time
complexity are noticeably higher than in the original implementation.

\section{Benchmarks}
\label{sec:benchmarks}

\subsection{Performance of binary trie implementations}
\label{sec:benchmarks-1}

We now evaluate the performance of the various binary trie
implementations mentioned in this paper, on synthetic benchmarks,
both for execution within Coq
or after extraction to OCaml.  The implementations and the benchmark
harness can be found in the companion development \cite{companion-dev}.
Here are the implementations:

\begin{description}
\item[\textbf{Original}:]
The original, nonextensional implementation described in
section~\ref{core}:
\begin{coq}
Inductive tree (A : Type) := Leaf : tree A | Node : tree A -> option A -> tree A -> tree A.
\end{coq}

\item[\textbf{Canonical}:]
The canonical binary tries of 
section~\ref{canonical}, based on space-efficient
representation:
\begin{coq}
Inductive tree' (A: Type) : Type := 
 | Node001: tree' A -> tree' A
 | Node010: A -> tree' A
 | Node011: A -> tree' A -> tree' A
 | Node100: tree' A -> tree' A
 | Node101: tree' A -> tree' A -> tree' A
 | Node110: tree' A -> A -> tree' A
 | Node111: tree' A -> A -> tree' A -> tree' A.
Inductive tree (A: Type) : Type := 
  | Empty : tree A
  | Nodes: tree' A -> tree A.
\end{coq}

\item[\textbf{Node01}:]
A nonextensional implementation that avoids storing option values in
nodes, using two constructors \lst{Node0} (node carrying no value) and
\lst{Node1} (node carrying a value) instead:
\begin{coq}
Inductive tree (A : Type) := 
  | Leaf : tree A
  | Node0 : tree A -> tree A -> tree A.
  | Node1 : tree A -> A -> tree A -> tree A.
\end{coq}
This is a middle point between the Original and the Canonical
implementations, both in terms of space efficiency and in terms of
complexity of implementation.

\item[\textbf{Sigma}:]
An extensional implementation built on top of the Original
implementation using a subset type to guarantee well-formedness of
trees, as in section~\ref{subset-types}:
\begin{coq}
Inductive tree (A : Type) := Leaf : tree A | Node : tree A -> option A -> tree A -> tree A.
Definition t (A: Type) : Type := { m: tree A | wf m }.
\end{coq}

\item[\textbf{GADT}:]
An extensional implementation based on an indexed type, also known as
a GADT (generalized algebraic data type), as described in
section~\ref{GADT}:
\begin{coq}
Inductive tree (A : Type) : kind -> Type :=
  | Leaf : tree A Empty
  | Node : forall (kl kr: kind) (l: tree A kl) (o: option A) (r: tree A kr),
                  not_trivially_empty kl o kr -> tree A Nonempty.
Inductive t (A: Type) : Type :=
  | Tree : forall (k: kind), tree A k -> t A.
\end{coq}

\item[\textbf{Patricia}:]
A nonextensional implementation based on the Patricia binary trees
from \emph{Functional Algorithms, Verified!}, 
\cite[\S{}12.3]{nipkow21:fav}.  Each node contains a prefix (a list of bits, represented as
a positive integer) that must be matched before reaching the node,
thus compressing skinny branches.
\begin{coq}
Inductive tree (A : Type) : Type :=
  | Leaf : tree A
  | Node : positive -> tree A -> option A -> tree A -> tree A.
\end{coq}

\item[\textbf{AVL}:]
The implementation of finite maps by Filliâtre and Letouzey 
\cite{DBLP:conf/esop/FilliatreL04}, using AVL balanced binary search trees.
It is much more versatile than our binary tries, as it supports keys
of any type equipped with a decidable total ordering, not just
positive numbers.  We include this implementation in the comparison
because it is the most efficient finite map data structure provided by
the Coq standard library.

\item[\textbf{Red-Black}:]
Another generic implementation of finite maps, using red-black
balanced binary search trees, developed by Letouzey \cite{mmaps}.
It supports keys of any type equipped with a decidable total ordering, 
like the AVL implementation, but performs better in general.
\end{description}

We compare the performance of these implementations using three benchmarks:
\begin{description}
\item[\textbf{Dense}:] The keys 1 to 2048 are successively inserted in a trie,
  then every key is looked up.  This produces a dense, perfectly
  balanced binary trie.  This is representative of how binary tries
  are used in the back-end of the CompCert compiler to represent
  control-flow graphs, in particular.

\item[\textbf{Sparse}:] As in the Dense test, we insert then look up a number of
  keys.  The keys correspond to words randomly chosen from an English
  dictionary, then converted to positive numbers by viewing the
  words as strings of bits, with 8 bits per character and an ASCII encoding.
  We have 5064 words with length ranging from 1 to 18 characters
  (average 8 characters).  This test produces sparse, poorly balanced
  tries with long, nearly empty branches.  It is representative of
  some uses of binary tries in the VST infrastructure, where keys are
  encodings of program identifiers.  The static analyses of the
  CompCert back-end also use sparse tries intensively.

\item[\textbf{Repeated}:] We insert the same seven keys (1 to 7) a million times
  in a trie.  The purpose of this test is to check that the trie
  remains small (7 nodes) and does not grow at each insertion.  It is
  representative of the use of binary tries as environments in a
  reference interpreter: there are few variables in the program being
  interpreted, but they are assigned many times.
\end{description}

For execution within Coq, we use \verb|vm_compute| as our primary
evaluation mechanism.  It implements call-by-value through compilation
to a virtual machine.  For comparison, we also give some numbers
obtained with the \verb|cbv| and \verb|lazy| mechanisms.  These are
interpreters that follow either call-by-value (\verb|cbv|) or
call-by-need (\verb|lazy|).  These
mechanisms are much slower than \verb|vm_compute|, hence we report
results for the Sparse benchmark only.  We also extract the benchmarks
to OCaml and compile them to native code using the \verb|ocamlopt|
compiler.  For OCaml executions, we report execution times, amount of
memory allocated (in bytes), and memory size of the data structure
being built (in bytes).  For executions within Coq, we were unable to
measure memory usage and only report execution times.  The
experimental results are listed in table~\ref{t:benchs}.

\def\separator#1{\cline{2-9} \\[-8pt] \multicolumn{9}{c}{\bf #1} \\}
\def\absolutenumber#1{\multicolumn{3}{l}{~#1}}

\begin{table}
\begin{center}
\begin{tabular}{@{}l@{}rrrrrrrr@{}}
 &	Original	 & Node01	 & Canonical	 & Sigma	 & GADT 	& Patricia     & AVL    & Red-Black
\\ 
\separator{Coq execution, \texttt{vm\_compute}, Sparse test}
\emph{Time in s} & \absolutenumber{1.78e-02}
\\ 
\emph{Relative time} & 100\% & 99\% & 78\% & 105\% & 155\% & 87\% & 1080\% & 778\%
\\ 
\separator{Coq execution, \texttt{vm\_compute}, Dense test}
\emph{Time in s} & \absolutenumber{1.10e-03}
\\ 
\emph{Relative time} & 100\% & 97\% & 100\% & 123\% & 152\% & 196\% & 2804\% & 960\%
\\ 
\separator{Coq execution, \texttt{vm\_compute}, Repeated test}
\emph{Time in s} & \absolutenumber{5.20e-01}
\\ 
\emph{Relative time} & 100\% & 97\% & 122\% & 506\% & 207\% & 238\% & 2577\% & 710\%
\\ 
\separator{Coq execution, \texttt{cbv}, Dense test}
\emph{Time in s} & \absolutenumber{1.97e-02}
\\ 
\emph{Relative time} & 100\% & 95\% & 100\% & 137\% & 219\% & 224\% & 3881\% & 1070\%
\\ 
\separator{Coq execution, \texttt{lazy}, Dense test}
\emph{Time in s} & \absolutenumber{3.56e-03}
\\ 
\emph{Relative time} & 100\% & 79\% & 94\% & 109\% & 118\% & 127\% & 11426\% & 4904\%
\\ 
\separator{Extraction to OCaml, Sparse test}
\emph{Time in s} & \absolutenumber{1.09e-02}
\\ 
\emph{Relative time} & 100\% & 92\% & 69\% & 101\% & 127\% & 49\% & 371\% & 339\%
\\ 
\emph{Allocated bytes} & \absolutenumber{1.07e+07}
\\ 
\emph{Relative alloc} & 100\% & 76\% & 56\% & 100\% & 151\% & 118\% & 97\% & 34\%
\\ 
\emph{Data size} & \absolutenumber{6.11e+06}
\\ 
\emph{Relative size} & 100\% & 75\% & 50\% & 100\% & 149\% & 22\% & 45\% & 44\%
\\ 
\separator{Extraction to OCaml, Dense test}
\emph{Time in s} & \absolutenumber{1.96e-04}
\\ 
\emph{Relative time} & 100\% & 102\% & 119\% & 101\% & 115\% & 198\% & 1724\% & 1102\%
\\ 
\emph{Allocated bytes} & \absolutenumber{6.88e+05}
\\ 
\emph{Relative alloc} & 100\% & 87\% & 87\% & 100\% & 156\% & 298\% & 568\% & 241\%
\\ 
\emph{Data size} & \absolutenumber{9.83e+04}
\\ 
\emph{Relative size} & 100\% & 67\% & 50\% & 100\% & 133\% & 117\% & 160\% & 133\%
\\ 
\separator{Extraction to OCaml, Repeated test}
\emph{Time in s} & \absolutenumber{3.67e-02}
\\ 
\emph{Relative time} & 100\% & 122\% & 155\% & 100\% & 146\% & 231\% & 1847\% & 409\%
\\ 
\emph{Allocated bytes} & \absolutenumber{6.56e+08}
\\ 
\emph{Relative alloc} & 100\% & 83\% & 90\% & 100\% & 168\% & 245\% & 271\% & 175\%

\end{tabular}
\end{center}
\caption{Performance figures for 6 implementations of binary trees and
for two reference implementations based on AVL or Red-Black balanced
trees.  The implementations and the benchmark tests are described in
the main text, section~\ref{sec:benchmarks-1}.  The tests were
executed on a single core of an AMD Ryzen 7 3700X processor running
Ubuntu Linux 20.4, using Coq version 8.15.1 and OCaml version 4.14.0
in ``no naked pointers'' mode.
All tests were repeated enough times so that the measured running time
is between 1 and 10 seconds.  Timing variations between multiple
measurements are below 5\%.  For the Repeated tests, the size of the data
structure is not reported, as it is very small.
}
\label{t:benchs}

\end{table}

We see that the Canonical implementation significantly improves on the
Original implementation for the Sparse test: Coq execution times
decrease by 22\%, OCaml execution times by 33\%, and memory allocations
and in-memory data size are halved.  For the Dense and Repeated tests,
memory allocation is reduced by about 12\%, and data size is halved,
but execution times increase a bit.

The Node01 implementation stands halfway between the Original and the
Canonical implementations in terms of OCaml execution times and memory
consumption.  Coq execution times show no improvement compared with
the Original implementation.  Combined with the lack of
extensionality, these results show that Node01 is not an interesting
alternative to Canonical.

The Sigma implementation performs exactly like the Original
implementation after extraction to OCaml.  This is unsurprising: the
subset type is erased during extraction, resulting in very similar
OCaml code for the Sigma and Original implementations.  As expected,
execution inside Coq is slowed down by the construction and
propagation of proof terms for the well-formedness invariant:
modestly so (5\% to 37\% depending on the Coq evaluation mechanism used)
for the Sparse and Dense tests, but dramatically so (a 5-times
increase) for the Repeated test.

The GADT implementation avoids the dramatic degradations of the Sigma
implementation, but runs 17\% to 100\% slower than the
Original implementation, both within Coq and after extraction to
OCaml.  The extra arguments to the \lst{Node} constructor increase memory
usage by 50\%.

The Patricia implementation is quite efficient on sparse maps, beating
all the other implementations after extraction to OCaml, and
second only to the Canonical implementation for evaluation within
Coq.  On dense maps, performance is worse than the Original and
Canonical implementations, owing to more complex algorithms, such as
the need to compute the common prefix between two lists of bits
represented as positive integers.

The AVL and Red-Black implementations perform poorly compared to the binary trie
implementations: between 3 and 114 times slower than the Original
implementation.  These are generic implementations of maps that, unlike
the binary trie implementations, do not take advantage of the
specific structure of keys as positive numbers.  Instead, these
implementations spends much time in rebalancing and in comparing
positive keys, a comparison that is quite slow since positive numbers
are isomorphic to lists of bits.  The Red-Black implementation is
always more efficient than the AVL implementation, by a factor of up to~4.

\subsection{Comparison with other dictionary data structures}
\label{sec:benchmarks-2}

CompCert and VST occasionally use binary tries as dictionaries, that
is, finite maps indexed by character strings instead of by positive
numbers.  This is achieved by a simple encoding of strings as bit
sequences represented as positive numbers.  It is interesting to compare
the performance of these dictionaries derived from binary tries with the
performance of other implementations of dictionaries.
Table~\ref{t:benchs2} shows some measurements for the following 4
dictionary implementations:

\def\separator#1{\cline{2-5} \\[-8pt] & \multicolumn{4}{c}{\bf #1} \\}
\def\absolutenumber#1{\multicolumn{3}{l}{#1}}

\begin{table}

\begin{center}
\begin{tabular}{lrrrr}
	& AVL	& Red-Black 	& Char-Trie	& Canonical
\\ 
\separator{Coq execution, \texttt{vm\_compute}}
\emph{Time in s} & \absolutenumber{1.86e-01}
\\ 
\emph{Relative time} & 100\% & 31\% & 195\% & 14\%
\\ 
\separator{Extraction to OCaml}
\emph{Time in s} & \absolutenumber{7.02e-03}
\\ 
\emph{Relative time} & 100\% & 35\% & 62\% & 154\%
\\ 
\emph{Allocated bytes} & \absolutenumber{1.04e+07}
\\ 
\emph{Relative alloc} & 100\% & 35\% & 84\% & 159\%
\\ 
\emph{Data size} & \absolutenumber{8.55e+05}
\\ 
\emph{Relative size} & 100\% & 90\% & 170\% & 331\%

\end{tabular}
\end{center}
\caption{Performance figures for 4 implementations of the dictionary data structure. The implementations and the benchmark tests are described in the main text, section~\ref{sec:benchmarks-2}.  The measurements use same the methodology and hardware and software configuration as in Table~\ref{t:benchs}.}
\label{t:benchs2}

\end{table}

\begin{description}
\item[\textbf{Canonical}]: This is the better of the
  binary trie implementations used in section~\ref{sec:benchmarks-1},
  composed with string to positive conversions.  In other words, the
  \lst{get} and \lst{set} operations take strings as keys, convert them to
  positive numbers on the fly, and invoke the \lst{get} and \lst{set}
  operations of the Canonical binary trie implementations.
\item[\textbf{AVL}:]  The AVL balanced binary search trees from Coq's
  standard library, using strings as keys.
\item[\textbf{Red-Black}:]  The Red-Black balanced binary search trees
  by Letouzey, using strings as keys.
\item[\textbf{Char-Trie}:]  A trie data structure that branches on
  characters, instead of on single bits like the binary tries.
  More precisely, each node of the trie carries a sparse, sorted association list
  mapping the next character to the corresponding sub-trie.
\end{description}

The benchmark used in table~\ref{t:benchs2} is similar to the Sparse
benchmark from section~\ref{sec:benchmarks-1}: 5064 words randomly chosen
from an English dictionary are inserted in the data structure, then
looked up.

For executions within Coq, the implementation based on canonical
binary tries is 2 to 14 times faster than the AVL, Red-Black and
Char-Trie implementations,
despite the overhead of converting strings to positive numbers at each
operation.  This can be explained by the way Coq represents strings: a
string is isomorphic to a list of characters, each character being
isomorphic to a 8-tuple of Booleans.  This makes comparisons between
characters and comparisons between strings rather expensive.  The AVL
and Red-Black implementations perform many string comparisons, and the
Char-Trie implementation performs many character comparisons.

After extraction to OCaml, Coq's characters are mapped to OCaml's
characters, which are small machine integers that can be compared very
efficiently.  Strings remain as lists of small integers.
Consequently, the AVL, Red-Black and Char-Trie implementations run
significantly faster than the binary trie-based implementation, up to
4 times faster for Red-Black.  The performance of the
dictionaries based on binary tries remains acceptable, given the
simplicity of the implementation, and could be improved by ``fusing''
the string-to-positive conversion and the recursive tree traversals,
avoiding the construction of positive numbers.

\section{Integration in CompCert}
\label{sec:CompCert}

\def\separator#1{\cline{2-6} \\[-8pt] \multicolumn{6}{c}{\bf #1} \\}

\begin{table}
\begin{center}
\begin{tabular}{lrrrrr}
& \multicolumn{1}{c}{Lines}
& \multicolumn{2}{c}{Time}
& \multicolumn{2}{c}{Memory usage} \\
& of code
& for 3.9 & 3.10 change
& for 3.9 & 3.10 change \\
\separator{Compilation}
Raytracer & 2879    &  0.42s   &  +2.4\% & 33 MiB   & 0.0\%
\\
Spass     & 82111   &  17.9s   &  +1.0\% & 96 MiB   & -32.3\%
\\
Unit1     & 19755   &  108.7s  &  -4.3\% & 2.1 GiB  & -5.9\%
\\
Unit2     & 36709   &  73.9s   &  -7.8\% & 1.5 GiB  & -8.6\%
\\
Unit3     & 118952  &  208.4s  &  -7.5\% & 5.4 GiB  & -17.1\%
\\
\separator{Interpretation}
Fib       & 19      & 12.8s    &  +2.2\% & 2.2 GiB  & -7.0\%
\\
Qsort     & 50      & 53.0s    &  +1.4\% & 2.8 GiB  & -9.1\%
\end{tabular}
\end{center}

\caption{Impact of the change to canonical binary tries on CompCert compilation
  and interpretation times and memory usage.  We used CompCert
  versions 3.9 and 3.10, built with Coq version 8.13.2 and OCaml
  version 4.12.  Measurements were taken on a single core of an AMD
  Ryzen 7 3700X processor running Ubuntu Linux 20.4.}
\label{t:benchs3}

\end{table}

The canonical binary tries described in this paper were integrated in
version 3.10 of the CompCert verified C compiler, replacing the
original, non-extensional tries that had been used since the beginning
of CompCert.  The implementation follows exactly the approach described in
this paper and its companion Coq development, but provides more
operations, such as mapping or folding a function over a trie,
converting a trie to a list of (key, value) pairs, and comparing two
tries.  Moreover, a variant of the \lst{combine} operation that improves
in-memory sharing between its result and its two arguments is
provided to reduce the memory usage of dataflow analyses.

The implementation of CompCert makes heavy use of binary tries to
represent several kinds of finite maps.  Control-flow
graphs are represented as functional arrays implemented by dense
binary tries.  Most dataflow analyses use finite maps from
pseudoregisters to an abstract domain.  These maps are represented by
sparse binary tries, omitting the pseudoregisters that are associated
with $\top$.  Finally, CompCert also provides a reference interpreter
that animates its formal C semantics.  During interpretation, memory
states are represented by two levels of binary tries, as mappings from
block identifier and block offset to byte contents
\cite[chap.~32]{appel14:plcc}.

We measured compilation and interpretation times and memory usage for
a handful of C programs, comparing CompCert 3.9, the last version that
uses the original implementation of binary tries, with CompCert 3.10,
the first version that uses the canonical binary tries.  (There are
other changes between 3.9 and 3.10, but they appear to have no
measurable impact on compilation and interpretation times.)  The
results are summarized in table~\ref{t:benchs3}.

For the compilation tests, we have five programs: Raytracer and
Spass, which are hand-written, and Unit1, Unit2, and Unit3, which are
automatically-generated unit tests with a peculiarity: all the tests
are put in a single function, resulting in huge functions that cause
high compilation times and memory usage.  For the interpretation
tests, we use two tiny C programs: the Fibonacci function and the
Quicksort algorithm.

As table~\ref{t:benchs3} shows, the switch to canonical binary tries
decreases memory usage of compilation and interpretation by up to
30\%.  This was to be expected, given that binary tries are used a lot
and the canonical kind has a more compact in-memory representation.
Compilation and interpretation times increase slightly, by up to
2.4\%, except for the ``one huge function'' unit tests, where
compilation times decrease noticeably, by up to 7.8\%.

For finer analysis, we profiled two of the compilations using the
Linux \texttt{perf} tool, focusing on the functions that account for
more than 0.5\% of the total compilation time.  For the Spass test,
binary tries operations account for 14\% of the CompCert 3.9 compilation
time, and slow down by about 20\% when moving to 3.10.  (This is
consistent with the 20\% slowdown on dense maps after extraction to
OCaml reported in table~\ref{t:benchs}.)  However, memory management
(allocation and GC) account for 25\% of the CompCert 3.9 time, and
speed up by 6\% in 3.10, owing to the reduced memory usage.  The two
effects almost cancel each other, resulting in a small slowdown
overall.  For the Unit3 test, binary tries operations account for 26\%
of the 3.9 compilation time, and take about the same time in 3.10,
while memory management accounts for 44\% of the 3.9
compilation time, and decrease by 12\% in 3.10 owing to the reduced
memory usage.  The result is a clear decrease in compilation times.

\section{Efficiency improvements in VST}
\label{sec:impactVST}

The Verified Software Toolchain (VST) 
~\cite{appel11:esop}
comprises a program logic
(Verifiable C) for the C programming language,
proved sound in Coq with respect to the operational semantics
of CompCert Clight, and a proof automation system (VST-Floyd)
\cite{vst-floyd}
to assist users in applying the program logic to their program.
Like CompCert, VST heavily uses PTrees for many purposes:
mapping function-names to function-bodies,
function-names to their specifications,
global variable names to their types,
local variable names to their types,
local variable names to their symbolic contents,
and so on.

Unlike CompCert, which runs in extracted OCaml, VST runs inside Coq's Gallina logic,
because it must manipulate proof goals and proof terms,
and even the PTrees must map names to values that contain fragments of
Coq terms and proofs.  Therefore, unlike in extracted OCaml where propositions
and proofs-of-propositions have been erased, computations inside VST
are unerased.  This means that the sigma-type representation would be
much less efficient in VST than it would be in CompCert.\footnote{In table~\ref{t:benchs}, performance of PTrees in VST could be predicted in row ``Coq execution, vm\_compute, Sparse test,''
column Sigma versus columns Original and Canonical;
whereas PTree performance in CompCert could be predicted in row ``Extraction to OCaml, Sparse test'', same columns.}

A recent enhancement to VST is \emph{Verified Software Units}, which allows
modular verification of modular C programs  \cite{beringer21:vsu}.
The VSU system uses PTrees even more heavily, in reasoning about linkage of
programs, of specifications, and of proofs.
Furthermore, in the VSU system it is often necessary to compare two PTrees for
equivalence, which is much faster to do in a data structure  with extensionality.

Furthermore, VST theorem statements (and intermediate proof goals)
often contain concrete PTrees, mapping C program identifiers to
various program specifications and AST terms.  A more compact representation
of these PTrees would reduce the proof-checking burden on the Coq kernel.

So it would not be surprising if a faster, more memory-efficient, extensional representation of PTrees would significantly improve performance of the VST interactive verifier.
And indeed, we are not suprised.  VST/VSU verifications run 10\% faster \emph{overall} with Canonical Binary Tries.

\paragraph{Measurement method.} We compared VST 2.8 (with original PTrees) and VST 2.9 (with canonical PTrees).  In release 2.9 there were no other changes that would significantly impact performance.  We benchmarked the ``pile'' benchmark \cite[Fig.~1]{BeringerAppel:FMSD21} as adapted to VSUs\footnote{This Coq verification is at \url{https://github.com/PrincetonUniversity/VST/releases/tag/v2.9.1} in
subdirectory progs/VSUpile.}
We measured performance of all the \emph{verifications} (\lstinline{verif_*.v});
we omitted specification files, which are not generally time-consuming and did not show
significant improvement.  Checking those Coq proofs\footnote{%
\lstinline{verif_stdlib},
\lstinline{verif_pile},
\lstinline{verif_onepile},
\lstinline{verif_apile},
\lstinline{verif_triang},
\lstinline{verif_core},
\lstinline{verif_main}}
totaled 105 seconds with original PTrees and 95.5 seconds with canonical PTrees\footnote{%
Measurements were performed on a Lenovo t440p laptop
with Intel Core i7-4810MQ @ 2.8GHz with 32GB memory, using Coq 8.14/8.15.1 in 32-bit mode with
virtual-memory limit of 1GB.  The Clight abstract-syntax-tree files were produced by
\lstinline{clightgen} in its default \lst{canonical-idents} mode.},
an improvement of 10\%.  

\section{Conclusion}

Our quest for finite maps that enjoy the extensionality property led
us to a new data structure, \emph{canonical binary tries,}
that performs very well in many contexts: extracted to OCaml and compiled
to native code, or represented in Coq proofs.  
In OCaml code, canonical binary trees reduce memory usage
significantly and improve execution times somewhat compared with simple
binary tries.  When keys are character strings, other implementations
such as balanced binary trees are faster, but they require
primitive integers, whose associated implementations would
then have to be trusted (or proved).

In functional programs written in Coq,
canonical binary tries are also significantly faster than
simple binary tries.  But when (concrete) tries must
appear in Coq theorem statements and in Coq proofs,
canonical tries offer additional benefits:
proof-checking efficiency (smaller theorem statements, smaller proofs)
and theorem simplicity (the ability to reason by Leibnitz equality
instead of equivalence relations).

At first, we were worried about the increase in the number of cases
for function definitions and for proofs, compared with the original
binary tries where there are only two cases to consider.  We demonstrated
two approaches to synthesizing large case analyses instead of writing
them by hand: one is based on Coq's tactic language and lets us
fluidly combine synthesis-time case analysis,
run-time case analysis, and synthesis-time symbolic analysis;
the other is based on custom constructors and induction principles
that lets us view the new data structure as the original, two-case
data structure, combined with partial evaluation within Coq to
eliminate the run-time overhead of the view.  More generally, Coq
offers powerful metaprogramming facilities that have great potential
to facilitate not just the verification but also the implementation of
complex data structures.

Besides binary tries, are there other useful data structures that
admit canonical representations?  And are these canonical
representations always more memory-efficient than other
representations?  We have no general answers to these questions, but
we can offer two other examples of canonical representations.

The first example is the clever representations of binary numbers used
in Coq (the type \lst{positive} of positive integers mentioned in 
section~\ref{core}) and in HOL, where natural numbers can be represented by
the constant \lst{ZERO} and two constructors \lst{BIT1} and \lst{BIT2}, with
\lstinline{BIT1($p$)} denoting $2p+1$ and \lstinline{BIT2($p$)}
denoting $2p+2$.
These representations are canonical, unlike the obvious representation
as lists of bits, and use less memory space, since the constructors \lst{xO},
\lst{xI}, \lst{BIT1}, \lst{BIT2} have one argument only, instead of the two
arguments of the ``cons'' list constructor.

For a less familiar example, consider sets of positive numbers.  The
usual representation of sets as lists
is neither canonical nor extensional: the set $\{1,2\}$ has several
representations, $[1,2]$ or $[2,1]$ or $[1,1,2,1,2]$.  We can recover
extensionality by using a subset type \lst{\{ l : list positive | sorted l \}},
where the \lst{sorted} predicate ensures that the list is increasing and
without repetitions.  However, a canonical representation exists, as
the list of (positive) differences from one set element to the next
greater element.  For instance, the set $\{1,4,9,11\}$ is uniquely
represented by the list $[1, 4-1, 9-4, 11-9]$, that is, $[1,3,5,2]$.
This encoding is not only canonical, but also slighty more memory
efficient than the usual sorted list representation, as the numbers
stored in the list of differences are smaller than those stored in the
sorted list.

Can we play similar tricks for more complex data structures?  We leave
this question for future work.

\bibliographystyle{spmpsci}
\bibliography{appel,biblio}

\end{document}